# Calculer le bilan Carbone de votre parc informatique avec EcoDiag, un service EcoInfo

**Béatrice Montbroussous**
GATE – UMR5824 / CNRS – RESINFO, EcoInfo

**Françoise Berthoud**
EcoInfo – GDS3524 / CNRS, EcoInfo

**Gabrielle Feltin**
GRICAD – UMS3758 / CNRS, EcoInfo

**Gabriel Moreau**
LEGI – UMR5519 / CNRS, EcoInfo

**Jonathan Schaeffer**
OSUG – UMS832 / CNRS, EcoInfo

## Résumé

*Vous entendez beaucoup parler des problématiques environnementales, vous avez peut-être initié des attitudes positives personnelles, et vous vous interrogez sur l'**impact environnemental du numérique** de votre **milieu professionnel … EcoDiag** est là pour vous !*

*Le GDS EcoInfo du CNRS a fait le constat que l'évaluation des **impacts environnementaux du numérique** est **complexe** et peut décourager les meilleures volontés, c'est pourquoi nous vous proposons une **méthode simple et efficace** pour estimer le bilan carbone de votre parc au travers de ce nouveau service basé sur les connaissances que nous avons engrangées. Nous avons choisi un indicateur compréhensible par tous : les émissions de **$CO_2e$ (Équivalent CO2)**. À partir d'un inventaire des matériels informatiques de l'unité, notre méthodologie et notre expertise vous permettront d'établir une **estimation globale des émissions de GES (Gaz à Effet de Serre)**. Cette évaluation sera accompagnée de préconisations et de ressources (fiches, guide, matériel) pouvant êtres utilisées dans le cadre d'un bilan carbone de votre structure ou être intégrée dans un rapport de type Hcéres. L'apport de conseils vous permettra d'afficher une politique de réduction des GES générés par le numérique dans les années suivantes !*

## Mots-clefs

*Impact environnemental, Bilan carbone, Bilan matériaux, Évaluation, Diagnostic, Parc informatique, Inventaire, Amélioration continue, Bonnes pratiques*

# 1 Introduction

L'année 2019 a été, dans la société française, dans les médias et dans nos institutions, marquée par une prise de conscience accrue des enjeux climatiques et de la pression de nos activités humaines sur les écosystèmes.

Le milieu de l'enseignement supérieur et de la recherche (ESR) n'est pas en reste : Flygskam [1], critère environnemental de l'Hcéres [2], Labo1point5 [3], initiatives au sein des universités et des instituts.

Depuis plus de dix ans, le GDS EcoInfo construit une offre de service pour la communauté ESR visant à réduire les impacts environnementaux des activités numériques. En effet, les technologies du numérique représentent aujourd'hui 3,7 % des émissions de gaz à effet de serre, et connaissent une croissance annuelle évaluée à 9 % [4]. Les impacts environnementaux liés au numérique sont importants dans toutes les phases du cycle de vie des équipements impliqués : extraction de matières premières non renouvelables, assemblage des matériaux, transports, usage et fin de vie.

Là aussi, dans notre communauté, la prise de conscience émerge. Mais s'attaquer au bilan carbone de l'informatique, c'est se heurter d'emblée à des difficultés importantes : obtenir des informations, évaluer leur fiabilité, croiser des études contradictoires, et parfois se heurter à un manque de données.

C'est pourquoi le GDS EcoInfo a mis en place un nouveau service afin de faciliter la mise en place d'un bilan carbone et d'une démarche d'amélioration continue à l'échelle d'un laboratoire.

Nous présentons dans cet article ce nouveau service EcoDiag : l'objectif, la méthodologie employée, l'efficacité et les limites de la méthode, ainsi qu'un retour d'expérience d'une mise en œuvre expérimentale. Enfin, nous présentons en quoi ce service représente une avancée, et terminons par quelques préconisations générales.

# 2 Objectif du service EcoDiag

L'objectif du nouveau service EcoDiag proposé par EcoInfo est de fournir aux informaticiens de notre communauté une méthode simple et efficace pour évaluer les émissions de Gaz à Effet de Serre (GES) de l'activité numérique au sein d'une unité.

À l'issue de la démarche EcoDiag, un gestionnaire de parc disposera d'un indicateur ramené en équivalents CO2, sur le périmètre considéré et sur une période d'un an. Pour exploiter correctement cet indicateur, il conviendra de renouveler l'exercice chaque année afin d'évaluer l'efficacité des mesures qui ont été mises en œuvre pour la réduction des émissions de GES liés à l'informatique.

La démarche proposée ici s'attache à prendre en compte la plus grande part possible des flux nécessaires aux activités numériques de l'unité, dans une approche de type Cycle de Vie. Ainsi les phases extraction des métaux, fabrication et de transport des équipements sont prises en compte.

Dans le secteur des Technologies de l'Information et de la Communication, il est particulièrement difficile de trouver des données, à cause de :

— la forte évolution technologique rendant les évaluations rapidement obsolètes dans des proportions difficiles à estimer ;

- la complexité des équipements et le multi-régionalisme de leur fabrication, la volatilité des sous-traitants et leur nombre ;
- la disponibilité des informations.

Du fait de cette complexité, les chiffres et estimations montrent rapidement leurs limites lorsqu'ils sont considérés de manière isolée. Cependant, ils prennent tout leur sens pour réaliser des comparaisons et orienter les choix vers l'option la moins impactante, ou pour évaluer l'évolution dans le temps de l'indicateur.

EcoDiag se présente comme un référentiel d'émissions de GES pour chaque type de matériel du périmètre considéré, permettant aux utilisateurs d'estimer l'impact global sur leur périmètre, mais aussi d'orienter leurs choix d'achat ou d'usage pour réduire ces impacts.

## 3 Méthodologie

Face à la diversité des situations, entre les recherches qui ont un usage « bureautique » du numérique et des approches de type simulation numérique ou celles qui manipulent de très gros volumes de données, face à l'évolution des technologies, le service propose des données qui sont suffisamment segmentées pour que chacun y trouve son compte et surtout des données qui seront mises à jour régulièrement.

Ce bilan de GES n'a pas pour vocation de comparer des organisations entre elles, mais il constitue un moyen d'évaluer le résultat des actions menées par chaque organisation relativement aux résultats des années précédentes.

### 3.1 Fabrication, usage et fin de vie

Le bilan proposé par EcoDiag tient compte des phases du cycle de vie des matériels. En particulier, pour les impacts liés à la fabrication, EcoDiag ne tient compte que des équipements acquis dans l'année en cours. Nous valorisons ainsi les unités qui font le choix de la limitation des achats.

Pour la comptabilisation des DEEE (Déchets Électriques et Électroniques), EcoDiag intègre une évaluation en équivalent CO2 pour l'équipement considéré.

### 3.2 Périmètre

La norme ISO 14064-1 [5] prévoit que la réalisation d'un bilan GES d'une organisation nécessite de définir le périmètre organisationnel, c'est-à-dire les installations ou les sites concernés.

Ainsi en fonction des situations, le périmètre organisationnel peut recouvrir plusieurs sites, tout ou partie des activités de la structure, un service donné, *etc*. Le point important étant que le périmètre doit être très clairement explicité.

Le périmètre opérationnel correspond aux catégories et postes d'émissions liées aux activités du périmètre organisationnel. Les principales normes et méthodes internationales définissent trois catégories d'émissions. Nous présentons ci-dessous les principaux postes d'émission correspondant.

### 3.2.1 Émissions directes de GES (SCOPE 1)

Les émissions directes provenant des équipements fixes ou mobiles situées à l'intérieur du périmètre organisationnel (détenus ou contrôlés par la structure) sont :

- les émissions directes fugitives (fuites de fluides frigorigènes dans les climatiseurs des salles serveurs).

### 3.2.2 Émissions à énergies indirectes (SCOPE 2)

Les émissions indirectes liées à la production d'électricité, de chaleur ou de vapeur importée pour les activités de l'organisation sont :

- les émissions indirectes liées à la consommation d'électricité ;
- les émissions indirectes liées à la consommation de froid.

### 3.2.3 Autres émissions indirectes (SCOPE 3)

Les autres émissions de GES indirectement générées par les activités de la structure et qui contribuent à la chaîne de valeur sont :

- les émissions liées à l'énergie non incluse dans les deux autres catégories (extraction, production et transport des combustibles consommés lors de la production d'électricité, de vapeur, de chaleur et de froid consommée par l'organisation) ;
- les achats de produits ou de services (sous-traitance et extraction, production des intrants matériels et immatériels de la structure) ;
- les immobilisations de biens (extraction et production des biens corporels et incorporels immobilisés par la structure) ;
- les déchets (transport et traitement des déchets de la structure) ;
- le transport de marchandise amont ;
- les déplacements professionnels ;
- le déplacement domicile-travail.

> EcoDiag s'applique aux équipements numériques matériels et immatériels sur les SCOPE 1 à 3.

## 3.3 Détail des équipements concernés

### 3.3.1 Le parc bureautique

Dans le parc bureautique, nous prenons en compte les équipements suivants, qu'ils soient en usage (scope 2 et 3) ou simplement stockés (seulement scope 3 dans ce cas) :

- les ordinateurs fixes et portables ;
- les tablettes électroniques ;
- les périphériques standards : écrans, claviers, souris ;

- les imprimantes de bureau ;

- les périphériques de stockage : clés USB, disques durs externes.

#### 3.3.2 La téléphonie

La téléphonie comprend les périphériques pour les téléphones IP et les téléphones mobiles professionnels.

#### 3.3.3 Les salles serveurs

Une salle serveur est un local permettant d'exploiter des serveurs informatiques. Par cette définition, une salle serveur peut être très petite, ne comportant que quelques stations de travail, héberger un ou plusieurs racks de serveurs, disposer d'une climatisation ou non.

Au sein d'une salle serveur, nous considérons :

- à la fois, les scopes 2 et 3 pour les serveurs et les stations de travail, les équipements réseaux, les baies de stockage ;

- seulement les scopes 1 et 2 pour les climatiseurs ;

- et le scope 2 pour les onduleurs.

#### 3.3.4 Les équipements communs

Ce sont les équipements informatiques partagés au sein de la structure. Nous comptabilisons les vidéoprojecteurs, les équipements de visioconférence, les bornes Wi-Fi, les copieurs multifonctions (scope 2 et 3).

#### 3.3.5 Le calcul scientifique

Les campagnes de calcul, soumises sur des mésocentres ou centres nationaux sont comptabilisées (scope 2).

#### 3.3.6 Hors périmètre

Les éléments qui ne sont pas cités ci-dessus ne sont pas comptabilisés. Voici quelques exemples et la raison pour laquelle nous les excluons du périmètre :

- les serveurs mutualisés : La diversité des fonctionnements de mutualisation ne nous permet pas de proposer une règle générale permettant d'évaluer l'impact d'une partie prenante ;

- les hébergements extérieurs : comme pour les serveurs mutualisés, les services d'hébergements extérieurs sont trop variés pour que nous puissions proposer un calcul générique dans EcoDiag. Cependant, certains fournisseurs de services proposent une information environnementale et il est alors possible de les inclure.

- les câbles : faire l'inventaire des câbles et de leur longueur nous a semblé contre-productif dans le cadre de la méthode EcoDiag. L'impact environnemental des câbles est lié à la production de plastiques, à l'extraction de cuivre et des dopants (métaux spécifiques) nécessaires si la fibre optique est utilisée. Nous fournissons néanmoins des valeurs pour les câbles réseau classe 5 et les câbles HDMI.

### 3.4 À propos de la phase d'usage des équipements terminaux et serveurs

La phase d'usage concerne la consommation électrique des éléments pendant sa durée de vie. Pour les équipements utilisateurs nous considérons que tous les équipements fonctionnent 1607 h par an (le temps de travail à temps plein), même si un utilisateur en possède plusieurs ou ne les éteint pas hors du temps de travail.

Pour les serveurs et les stations de travail, nous considérons qu'ils fonctionnent 24h/24 toute l'année.

### 3.5 Fiabilité des sources

Lors de l'élaboration de la base de référence EcoDiag, nous avons étudié plusieurs sources qui ont mis en évidence des disparités importantes. Nous discutons dans cette section la fiabilité des différentes sources utilisées.

### 3.6 Pourquoi de telles incertitudes dans les valeurs ?

Pour plusieurs raisons :

— d'abord, parce qu'on ne sait pas estimer précisément depuis l'extraction des métaux jusqu'à l'assemblage final l'ensemble des GES émis par les différents acteurs de la chaîne. Les données sont des moyennes (par exemple moyenne des GES émis par l'extraction d'un kg de cuivre sur un grand nombre de mines), avec des incertitudes. À noter que les données de base sont presque toujours issues de la base EcoInvent [6];

— ensuite, parce que en général, les GES émis sont moyennées sur plusieurs modèles de la même série dans des configurations de base (2e niveau de moyenne), par exemple les équipements *Latitude Série 5xxx* ;

— enfin, parce que parfois on utilise des données d'un équipement proche.

> En conséquence, le résultat final donne un ordre de grandeur et non une valeur fine exacte.

### 3.7 Comment évaluer la fiabilité d'un résultat ?

Nous avons jugé de la fiabilité d'un résultat en tenant compte des éléments suivants :

— les commanditaires de l'étude sont-ils neutres ? Ont-ils des intérêts financiers liés aux résultats ? ;

— l'étude est-elle revue par des pairs, citée ? S'appuie-t-elle sur d'autres études sérieuses ?

Nous avons aussi jugé de la pertinence d'une évaluation en étudiant les hypothèses et les périmètres sur lesquels elle s'appuie.

### 3.8 Quelques exemples

#### 3.8.1 Le guide sectoriel TIC ADEME

Le bilan carbone tel que proposé par l'ADEME et notamment le guide « Réalisation d'un Bilan des émissions de gaz à effet de serre ; guide sectoriel : Technologies Numériques, Information et Communication » [7] est un document précieux sur lequel nous nous sommes appuyés. Cependant, les données sont souvent obsolètes et nous nous sommes attachés à les vérifier et à les mettre à jour pour EcoDiag.

#### 3.8.2 La base Carbone ADEME

La base Carbone de l'ADEME est la plus complète que nous ayons trouvée. Malheureusement sur nombre de produits, les données sont un peu anciennes avec des taux d'incertitude très importants. Aussi chaque fois que possible nous avons choisi une autre source plus récente (EcoInvent, données constructeurs…).

#### 3.8.3 Le livre blanc ADEME 2015

Ce rapport [8] intitulé « Consommation énergétique des équipements informatiques en milieu professionnel » sur les consommations électriques en phase d'usage a été commandité et publié par l'ADEME en 2015. Il fait une synthèse des consommations électriques de 50 entreprises et structures universitaires du sud-ouest de la France relevées entre 2012 et 2015. Le périmètre de cette étude nous a paru suffisamment proche de nos structures pour l'utiliser en première approximation. Les utilisateurs qui disposent de leurs propres données de consommation sont évidemment invités à les utiliser.

#### 3.8.4 Les fiches environnementales de Dell, Apple ou Seagate

Les constructeurs Dell et Seagate publient une fiche environnementale [9] pour chacun de leurs produits. Il s'agit d'une évaluation d'empreinte CO2 de chaque matériel détaillant la part de chaque phase du cycle de vie. Le constructeur s'appuie sur la méthode Product Attribute to Impact Algorithm (PAIA) [10] développée par le MIT. Compte tenu des hypothèses choisies par la méthode (comptabilisation d'un PUE théorique, équivalent carbone du kWh européen ou états-unien, durée d'usage et conditions d'usages), nous n'avons retenu de ces fiches que les données relatives à la fabrication et au transport des équipements. Nous avons appliqué la même démarche pour les fiches environnementales de Apple ou Seagate.

## 4 Retour d'expérience

Avant de proposer cette méthode et les ressources associées à toute notre communauté, nous avons souhaité la tester auprès d'un laboratoire volontaire.

### 4.1 Contexte

Le service informatique de l'UMR GATE à Lyon (Groupe d'Analyse et de Théorie Économique) s'est porté volontaire pour tester EcoDiag et faire un retour au groupe de travail. Cette unité possède un parc informatique de 150 postes clients, 35 serveurs dans une salle climatisée et ondulée, un copieur multifonction, quelques imprimantes individuelles et smartphones. Il a en charge la gestion du réseau, de la ToIP et du contrôle d'accès, avec les équipements ad-hoc. Ce parc recouvrant l'ensemble du périmètre proposé par EcoDiag, l'exercice a permis d'éprouver notre méthode de manière complète.

La méthode a été testée par la responsable du service informatique de l'UMR GATE.

### 4.2 Déroulement d'EcoDiag

*EcoInfo* : Qu'est-ce qui vous a motivé pour entreprendre la démarche EcoDiag ?

*GATE* : J'ai entrepris la démarche EcoDiag en raison d'un questionnement qui me bloquait dans mes prises de décision sur l'évolution matérielle du système d'information de mon unité. En effet, j'ai une baie de serveurs datant de 14 ans et je m'interroge sur la marche à suivre à savoir si je les remplace alors qu'ils fonctionnent encore et qu'ils sont en service, ou est-ce que je les garde alors qu'ils consomment probablement plus que des matériels plus récents ? N'ayant pas de compétence dans ce domaine mais étant préoccupée par cette problématique, je me suis tournée vers des personnes en capacité de m'aider.

*EcoInfo* : Quelles difficultés avez-vous rencontré pendant la démarche ? Comment les avez-vous surmontées ?

*GATE* : La plus grande difficulté rencontrée a été de fournir les consommations électriques des différents matériels, car je n'ai aucune mesure, et pas le temps d'en faire par quelque méthode que ce soit. J'ai espéré que EcoDiag me fournirait une solution de remplacement, et en effet à défaut de mes retours, les calculs se sont basés sur des informations théoriques ou des mesures externes. À mes yeux, ces résultats sont déjà très importants et utiles pour me faire un premier avis.

*EcoInfo* : Qu'allez-vous faire avec les résultats obtenus ? Qu'attendez-vous d'EcoInfo pour la suite ?

*GATE* : Les résultats obtenus vont me permettre de prendre de meilleures décisions sur les équipements à remplacer, et sur les futurs investissements à réaliser pour l'unité. Ils vont faire partie des arguments factuels que j'avancerai pour soutenir les choix que je présenterai à la direction. D'autre part, je compte aussi sur le regard extérieur de EcoInfo qui pourra être posé sur nos pratiques quotidiennes afin de les modifier et ainsi réduire l'impact du numérique de tous dans le laboratoire [11].

## 5 Discussion

### 5.1 Limitations d'EcoDiag

Nous avons bien conscience qu'il serait pertinent et important de dresser de la même façon un bilan matière des activités liées au numérique, mais nous ne disposons pas des ressources nécessaires (humaines notamment) pour proposer un tel indicateur.

### 5.2 Avancées et difficultés

La méthode EcoDiag met à la portée de tous la démarche d'un bilan carbone simplifié pour son parc informatique. Le travail de collecte et de compilation des sources de données, d'évaluation de la fiabilité des valeurs, a été réalisé par une équipe d'ingénieurs du GDS EcoInfo dans le but de diffuser un référentiel commun au sein de la communauté enseignement supérieur et recherche. En cela, EcoDiag est une grande avancée.

Cependant, la collecte des données par ceux qui se prêtent à l'exercice peut s'avérer complexe. Par exemple, la quantité de gaz réfrigérant injecté dans le groupe froid chaque année, la consommation électrique globale d'une salle serveur, sont souvent difficiles à obtenir.

Il ne faut pas que cette difficulté soit un frein pour engager une démarche EcoDiag. Nous considérons qu'il est préférable d'avoir une évaluation approximative permettant de mettre en place une démarche d'amélioration continue plutôt que de n'avoir aucune évaluation, faute de mesure suffisamment précise.

### 5.3 Préconisations générales

Dans un premier temps, il est important de disposer d'un inventaire suffisamment précis et à jour pour tout le périmètre couvert par EcoDiag. Le logiciel libre GLPI [12] est couramment utilisé dans notre communauté et nous le recommandons pour réaliser cet inventaire.

Nous recommandons, dès l'achat de l'équipement, d'indiquer dans l'inventaire sa consommation électrique moyenne ou son empreinte carbone. Cela facilitera grandement votre bilan EcoDiag futur.

Concernant les salles serveurs, si possible, mettre en place des systèmes de mesures au plus près de chaque matériel dès la construction, ou lors de rénovations. Si certains serveurs peuvent être interrogés pour connaître une consommation électrique précise [13], ce n'est pas le cas de beaucoup d'autres équipements comme les baies de stockage ou les routeurs. Pour une mesure de consommation précise, l'utilisation de PDU (Power Distribution Unit) communicants est la meilleure méthode. En l'absence de ces relevés, des valeurs théoriques sont proposées dans le cadre de la méthode.

> Nous recommandons de mettre en œuvre une sobriété numérique.
> Choisir les modèles au plus juste des besoins, en pensant à les faire durer le plus longtemps possible, éviter les doubles ou triples affichages, choisir les solutions de redondance seulement lorsqu'elles se justifient.

L'électricité en France métropolitaine est particulièrement peu carbonée. Nous proposons d'utiliser la valeur de la base de données EcoInvent [14] (0,119 kgCO2e) bien qu'elle corresponde à une valeur de pics de production. Cette valeur très basse relativement aux moyennes européennes ou mondiales est due à la forte proportion de nucléaire dans le mix énergétique français. Il est important de garder à l'esprit les autres problématiques que cela engendre et qui ne sont pas traduites en CO2 (l'impossibilité du démantèlement, le stockage des déchets, *etc*.). Donc, même si une économie électrique n'améliorera pas sensiblement le score EcoDiag, il reste important de faire baisser la demande en électricité [15]. Pour le parc informatique, les recommandations habituelles sur la veille des ordinateurs et des écrans restent tout à fait valables ; pour les datacentres, de nombreuses améliorations peuvent être apportées et EcoInfo propose de réaliser des audits spécifiques.

Dans le cas des calculs numériques, nous recommandons de sensibiliser les programmeurs et les chercheurs à l'optimisation, faire la promotion des langages de programmation sobres ou le mieux adapté aux usages (optimisations matérielles par exemple).

Nous recommandons aussi de publier les chiffres obtenus, afin de responsabiliser les utilisateurs qui, souvent sont de bonne volonté face aux enjeux climatiques.

Enfin, la méthode EcoDiag est conçue pour être réalisée en autonomie, mais EcoInfo peut vous accompagner pour sa mise en œuvre ou pour l'interprétation des résultats, ainsi que pour donner des recommandations pertinentes.

### 5.4 Après EcoDiag

Le GDS EcoInfo élabore actuellement une collection de bonnes pratiques qui peuvent être mises en œuvre dans la communauté ESR, et constitueront un bon moyen de dynamiser une démarche de sobriété numérique à la suite de l'évaluation EcoDiag. Ces bonnes pratiques seront publiées sur le site web d'EcoInfo [16].

## Annexe

Le tableau de référence de la méthode EcoDiag est disponible sur le site web d'EcoInfo.

## Bibliographie